\documentclass[a4paper, amsfonts, amssymb, amsmath, reprint, showkeys, nofootinbib, twoside, superscriptaddress]{revtex4-1}
\usepackage[english]{babel}
\usepackage[utf8]{inputenc}
\usepackage[colorinlistoftodos, color=green!40, prependcaption]{todonotes}
\usepackage{amsthm}
\usepackage{mathtools}
\usepackage{physics}
\usepackage{xcolor}
\usepackage{graphicx}
\usepackage[left=18mm,right=18mm,top=35mm,columnsep=15pt]{geometry} 
\usepackage{adjustbox}
\usepackage{placeins}
\usepackage[T1]{fontenc}
\usepackage{lipsum}
\usepackage{csquotes}
\usepackage{siunitx}
\usepackage[justification=centering]{caption}
\usepackage{subcaption}
\usepackage{comment}
\usepackage[pdftex, pdftitle={Article}, pdfauthor={Author}, colorlinks=true]{hyperref}
\begin{document}
\title{
Parameter scanning in a quantum-gravity-induced entanglement of masses (QGEM) experiment with electromagnetic screening
}

\author{Martine Schut}
    \thanks{Currently at: Centre for Quantum Technologies, National University of Singapore, 3 Science Drive 2, Singapore 117543}
    \affiliation{Van Swinderen Institute for Particle Physics and Gravity, University of Groningen, 9747AG Groningen, the Netherlands }
\author{Anupam Mazumdar }
    \affiliation{Van Swinderen Institute for Particle Physics and Gravity, University of Groningen, 9747AG Groningen, the Netherlands }


\begin{abstract}
Witnessing the quantum nature of spacetime is an exceptionally challenging task. However, the quantum gravity-induced entanglement of matter (QGEM) protocol proposes such a test by testing entanglement between adjacent matter-wave interferometers. 
One key obstacle to experimentally realising this protocol is the creation of a spatial quantum superposition with heavy masses. 
Initially, it was envisaged that the superposition size would have to be of order $200$ micron for a mass $10^{-14}$~kg (to obtain the entanglement phase of order unity when the centre of mass of the two interferometers are at a separation of $450$ microns).
The experimental design has since improved, e.g. by assuming that the two interferometers are separated by an electromagnetic screen, which helps bring the separation distance close to $35$ micron.
Here, we do parameter scans taking into account the electromagnetic screening, and we consider different geometrical setups; we show superpositions of at least a micron-size for mass $10^{-14}$~kg with a decoherence rate of order $10^{-3}$~Hz are required. 
\end{abstract}


\maketitle

\section{Introduction}\label{sec:intro}

Witnessing the quantum nature of spacetime in a lab is one of the profound questions that needs to be settled.  
There is now a protocol to test the quantum nature of gravity, as pointed out first in refs.~\cite{Bose:2017nin,ICTS}, see also ref.~\cite{Marletto:2017kzi}. 
The witness is based on a bonafide feature of quantum mechanics, known as an entanglement, which signifies the quantum correlation, which is very different from a classical correlation~\cite{Horodecki:2009zz}. 

The entanglement between two quantum systems, such as qubits, requires quantum interaction or a quantum mediator, which corroborates that local operation and classical communication (LOCC) cannot entangle the two quantum systems~\cite{Bennett:1996gf,Marshman:2019sne,Bose:2022uxe}. 
Hence, if spacetime is quantum, the two geometries with massive superpositions must be entangled via a quantum interaction mediated by a quanta, known as spin-2 graviton~\cite{Danielson:2021egj,Carney_2019,Carney23_nu,Biswas:2022qto,Marshman:2019sne,Bose:2022uxe,christodoulou2023locally,Vinckers:2023grv,chakraborty2023distinguishing,elahi2023probing,christodoulou2019possibility,Rufo:2024ulr}. 

The protocol given in ref.~\cite{Bose:2017nin} is known as the quantum gravity-induced entanglement of masses (QGEM). The authors confirmed that at a Newtonian level the gravitational entanglement can be established, see~\cite{Bose:2017nin,Bose:2022uxe,Danielson:2021egj,Carney_2019,Carney23_nu,christodoulou2019possibility,christodoulou2023locally,Vinckers:2023grv}. 
The gravitational potential and the Hamiltonian governing quantum gravity and its interaction are all operator-valued entities, therefore, any change in the gravitational potential is an operator-valued entity, which leads to entangling the two quantum objects~\cite{Bose:2022uxe}.

The QGEM protocol can be used to probe many interesting physics, such as the quantum weak equivalence principle~\cite{Bose:2022czr}, testing beyond the standard model physics~\cite{Barker:2022mdz}, constraining theories with massive gravity~\cite{elahi2023probing}, particular class of quantum gravity theories~\cite{Vinckers:2023grv,chakraborty2023distinguishing}, and testing the quantum analogue of light bending experiment in the context of witnessing entanglement between matter and photon degrees of freedom~\cite{Biswas:2022qto}. The latter experiment is based on an optomechanical test of quantum gravity that will verify the spin-2 nature of the graviton as a mediator. This is because only spin-2 graviton gives the right bending of light as predicted by Einstein's general theory of relativity,~see~\cite{Scadron:2007qd}.

The QGEM protocol, as envisaged in ref.~\cite{Bose:2017nin}, is based on creating a macroscopic quantum superposition with two nanoparticles of mass $m\sim 10^{-14}-10^{-15}$~kg. 
In the initial work of ref.~\cite{Bose:2017nin}, the authors assumed $\Delta x = 250\,{\rm \mu m}$ (the spatial superposition width), while $d_\text{min} = 200\,{\rm \mu m}$ (the minimal distance between any two superposition instances) to get an entanglement phase of order ${\cal O}(1)$.
This minimal distance was initially introduced such that the gravitational interaction between two dielectric spheres dominates sufficiently over the Casimir-Polder interaction~\cite{Casimir:1947kzi} which can cause a photon-mediated entanglement.
However, since then, the experimental parameters, such as $d_{min}$ and the entanglement phase, and as a result, the necessary superposition size, have greatly improved.
This was done by considering changes to the setup, e.g., by changing the geometrical configuration to one known as the parallel setup (which for the above parameters has a larger entanglement generation rate), and by considering the PPT (positive partial transpose) instead of CHSH (Clauser–Horne–Shimony–Holt) witness~\cite{Chevalier:2020uvv,Tilly:2021qef}, which required a smaller effective entanglement phase.
Further work, given in refs.~\cite{vandeKamp:2020rqh,Schut:2023eux}, showed that the minimal distance $d_\text{min}$ could be reduced by introducing a (super)conducting plate~\footnote{We would like to mitigate the photon-induced entanglement between the two superposed masses. 
Hence, even for a neutral nanoparticle, we wish to suppress the dipole-dipole potential, Casimir-Polder potential, and induced magnetic dipole potentials between them. 
It is possible to suppress these interactions via a conducting screen. 
However, there will be screen-sphere potentials we need to consider; these will result in an attractive/repulsive force between the screen and the test mass and will cause dephasing.}, and even further in combination with diamagnetic trapping~\footnote{
Levitation of the nanoparticles, can be done via diamagnetic levitation with via fixed magnetic setup~\cite{Hsu:2016}, or via chip-based levitating setup~\cite{Elahi:2024dbb}, see for details~\cite{Schut:2023eux,Elahi:2024dbb}.
After considering all these image potentials due to the conducting screen, it is possible to bring $d_{min}$ from $200\,{\rm \mu m}$ to $35{\rm \mu m}$ in a trapped setup~\cite{Elahi:2024dbb}.}; see refs.~\cite{Schut:2023hsy,elahi2023probing}.
This all has allowed for the necessary superposition size to be much smaller, requiring $\Delta x \geq 10\,{\rm \mu m}$ and $d_\text{min}=35\,{\rm \mu m}$ for mass $m=10^{-15}$~kg.

The most challenging part of the QGEM protocol is creating a large spatial superposition, and the work on the experimental design that has managed to reduce this distance is, therefore, useful for eventual experimental realisation. 
Generally, for the QGEM protocol, we consider a nitrogen-vacancy centre (NVC) in diamond and use the Stern-Gerlach effect of the spin in a magnetic field gradient to create a spatial superposition~\cite{Wan16_GM,Scala13_GM,
Bose:2017nin,Pedernales:2020nmf,Marshman:2021wyk,Marshman:2018upe,Zhou:2022epb,Zhou:2022frl,Zhou:2022jug,Zhou:2024voj,Braccini:2024fey}. 
The diamond NVC is convenient because of its spin readout properties, which are used to construct the witness and find any spin-spin quantum correlations. 
As with any quantum experiment, the QGEM protocol is also sensitive to dephasing and decoherence~\cite{Toros:2020dbf, Rijavec:2020qxd,
vandeKamp:2020rqh,Schut:2021svd,Schut:2023eux,Schut:2023hsy,Fragolino:2023agd,Schut:2024lgp}.
For a review of decoherence, see~\cite{bassireview,ORI11_GM}.  

The previous analysis of the trapped and shielded parallel setup - which is the setup that currently allows for the smallest superposition size~\cite{Schut:2023hsy,Elahi:2024dbb} - was done with an approximation for the witness expectation value. In this paper, we give the expectation value directly from the definition of the smallest eigenvalue of the partial transpose of the density matrix (the PPT witness expectation value)~\cite{PeresPPT}, see sec.~\ref{sec:parallel}.
We additionally analyse previous setups such as the linear setup~\cite{Bose:2017nin} and the 3-qubit parallel setup~\cite{Schut:2021svd} with the newly found parameters of the trapped and shielded setup, see sec.~\ref{sec:comparison}.

To summarise, the current paper gives us a much better judgment of the superposition size in conjunction with the decoherence rate and the entanglement witness parameter, $\langle {\cal W}\rangle$ (defined below). 
As we will show, in the most optimistic scenarios, $\Delta x \sim {\cal O}(1)\,{\rm \mu m}$ can give a witness $\langle {\cal W}\rangle<0$ for a decoherence rate of $\gamma\sim 10^{-3}$~Hz, for mass $m\sim 10^{-14}$~kg and $d_{min}=35\,{\rm \mu m}$. This parameter space is the most favourable parameter the future experiments can target to witness the entanglement due to the quantum nature of gravity.

\section{Witness expectation value for parallel setup}\label{sec:parallel}

Here, we briefly recall the parallel setup, which is most commonly used in recent designs of the QGEM protocol. 
The initially separable state gives the wavefunction of the two nanoparticles with respective spin superpositions:
\begin{equation}\label{WF-0}
    \Psi(t=0) = \frac{1}{2} \left( \ket{\uparrow} + \ket{\downarrow} \right)_1 \otimes \left( \ket{\uparrow} + \ket{\downarrow} \right)_2 \,.
\end{equation}
After the experiment is over, meaning the one-loop interferometer is conducted, the combined state becomes non-separable and entangled:
\begin{align}\label{WF-1}
    \Psi(t=\tau) = \frac{e^{i\omega \tau}}{2} \Big(& \ket{\uparrow}_1\ket{\uparrow}_2 + e^{i\omega_1\tau} \ket{\downarrow}_1\ket{\uparrow}_2 \nonumber \\ &+ e^{i\omega_2\tau} \ket{\uparrow}_1\ket{\downarrow}_2 + \ket{\downarrow}_1 \ket{\uparrow}_2 \Big) \, .
\end{align}
where, the phase $\phi_i\equiv \omega_i \tau$ is given by the gravitational interaction: $\phi \sim \hat U \tau/\hbar$, where $\hat U = G m^2/\hat{r}$ denotes the gravitational potential.
Eq.~\eqref{WF-1} is general for a bipartite quantum system, but the geometry of the setup determines its phases.
For the setup shown in Fig.~\ref{fig:setup_par}, known as the `parallel setup', the entanglement rates $\omega_1, \omega_2$ in eq.~\eqref{WF-1} are~\footnote{Making the simplification that $\Delta x$ is not time-dependent, i.e. the creation/annihilation of the spatial superpositions takes a very short time compared to the experimental time $\tau$ during which $\Delta x$ is kept constant. In practice, the particular Stern-Gerlach type protocol would determine the time dependence of $\Delta x$.}:
\begin{align}
    \omega_1^\text{par} &= \omega_2^\text{par} = \frac{Gm^2}{\sqrt{d_\text{min}^2+\Delta x^2}} \frac{1}{\hbar} - \frac{Gm^2}{d_\text{min}} \frac{1}{\hbar} \, . \label{eq:phases}
\end{align}
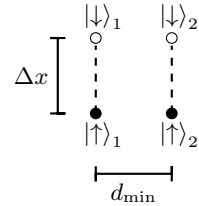
\begin{figure}[t]
\centering
\begin{tikzpicture}
\draw[black, dashed, thick] (0,-1) -- (0,-0.1);
\draw[black, dashed, thick] (1,-1) -- (1,-0.1);
\draw[black, thick] (0,-1.7) -- (0,-1.9) |- (0,-1.8) -- (1,-1.8) node[pos=0.5, below] {$d_\text{min}$} -| (1,-1.7) -- (1,-1.9);
\draw[black, thick] (-0.6,-1) -- (-0.4,-1) |- (-0.5,-1) -- (-0.5,0) node[pos=0.5, left] {$\Delta x$} -| (-0.6,0) -- (-0.4,0);
\filldraw[black]
(0,-1) circle (2pt) node[align=center, below] {\hspace{2mm}$\ket{\uparrow}_1$}
(1,-1) circle (2pt) node[align=center, below] {\hspace{2mm}$\ket{\uparrow}_2$};
\draw[black]
(0,0) circle (2pt) node[align=center, above] {\hspace{2mm}$\ket{\downarrow}_1$}
(1,0) circle (2pt) node[align=center, above] {\hspace{2mm}$\ket{\downarrow}_2$};
\end{tikzpicture}
    \caption{Two neutral test masses of mass $m$, labelled $1$ and $2$, in the parallel configuration. The superposition width is $\Delta x$, and the distance between the $\ket{\uparrow}$-states is $d_\text{min}$. 
    A superconducting screen can be placed in between the two superpositions to mitigate electromagnetic interactions such as dipoles and higher-order multipoles, 
    see~\cite{Schut:2023eux,Elahi:2024dbb}. However, the gravitational interaction between the two superpositions cannot be screened.}
    \label{fig:setup_par}
\end{figure}

The total entanglement phase is given by:
$ \tau (\omega_1+\omega_2)$.
Note that for our analysis, we will consider the maximum entanglement phase and ignore its time dependence. This makes sense as we are interested in the maximum entanglement phase accumulated in the experiment. 

The entanglement witness $\mathcal{W}$ is defined by the Positive Partial Transpose (PPT) witness~\cite{PeresPPT}, first introduced for the QGEM experiment by ref.~\cite{Chevalier:2020uvv}, and then implemented in refs.~\cite{Tilly:2021qef,Schut:2021svd}. 
The partial transpose of the tensor product of eigenvectors corresponds to the minimal eigenvalues of the partial transpose density matrix, the $\mathcal{W}$ is given by:
\begin{equation}
    \mathcal{W} = \left(\ket{\lambda_-^{T_2}}\bra{\lambda_-^{T_2}}\right)^{T_2} \, , \label{eq:ppt_def2}
\end{equation}
where $\lambda_-^{T_2}$ is the most negative eigenvalue of the partial transpose of the density matrix (denoted $\rho^{T_2}$). The expectation value of the PPT witness is given as:
\begin{equation}
    \langle \mathcal{W} \rangle 
    \equiv \Tr(\mathcal{W}\rho) = \Tr(\ket{\lambda_-^{T_2}}\bra{\lambda_-^{T_2}}\rho^{T_2}) = \lambda_-^{T_2} \, , \label{eq:ppt_def}
\end{equation}
In an experiment, we should include the effect of decoherence, as it is an essential part of our experiment. Hence, the density matrix of the wavefunction presented in Eq.~\eqref{WF-1} picks up decoherence terms given by: $\bra{i\,j}\rho\ket{i'\,j'} \rightarrow e^{-\gamma \tau (2 - \delta_{i,i'} - \delta_{j,j'})}$, where $\delta_{i,i'}$ and $\delta_{j,j'}$ are the Kronecker delta functions, and the total decoherence rate is given by $\gamma$.
The decoherence will lead to exponential damping in the density matrix; see ref.~\cite{Schlosshauer:2019ewh}.

At $t=\tau =0$, the density matrix is a pure state; this may be a challenging task experimentally.
However, for our analysis, we consider that creating such a pure initial state is possible~\footnote{In the future, we can easily generalise that to a thermal state by taking the thermal density matrix,see~\cite{Rizaldy:2024viw,Zhou:2024pdl} However, here, we will stick to a low-temperature limit, like the pure state case, by achieving motional ground-state cooling in the experiment similar to ~\cite{Deli__2020,Windey_2019, Piotrowski:2022qda, Tebbenjohanns_2020,Kamba:2023zoq} }. 
After taking the partial transpose of the density matrix, we find its eigenvalues:~\cite{Chevalier:2020uvv,Schut:2023eux}
\begin{align}
    \lambda_{1,2} &= \frac{1}{4} - \frac{1}{4} e^{-\gamma \tau} \left[ e^{-\gamma \tau} \mp 2 \sin(\frac{\omega_1\tau+\omega_2\tau}{2})\right] \, , \label{eq:witness_def} \\
    \lambda_{3,4} &= \frac{1}{4} + \frac{1}{4} e^{-\gamma \tau} \left[ e^{-\gamma \tau} \pm 2 \cos(\frac{\omega_1\tau+\omega_2\tau}{2}) \right]  \, 
    \label{eq:witness_def1} 
\end{align}
Equation~\eqref{eq:witness_def} gives the expectation value of the witness; this can be easily seen from the eigenvalues when taking $\gamma=0$, in which case $\lambda_1,\lambda_2\in[-\frac{1}{2},\frac{1}{2}]$ and $\lambda_3,\lambda_4\in[0,1]$.
A negative witness expectation value, $\Tr(\mathcal{W}\rho)<0$, indicates entanglement, while, for a 2-qubit system, a positive witness indicates no entanglement~\cite{Horodecki:2009zz,PeresPPT}.
Adding a non-zero decoherence rate increases the witness expectation value, making it less negative and thus making it harder to witness any entanglement.

\begin{figure*}
     \centering
     \begin{subfigure}[b]{\textwidth}
        \centering
        \includegraphics[width=\linewidth]{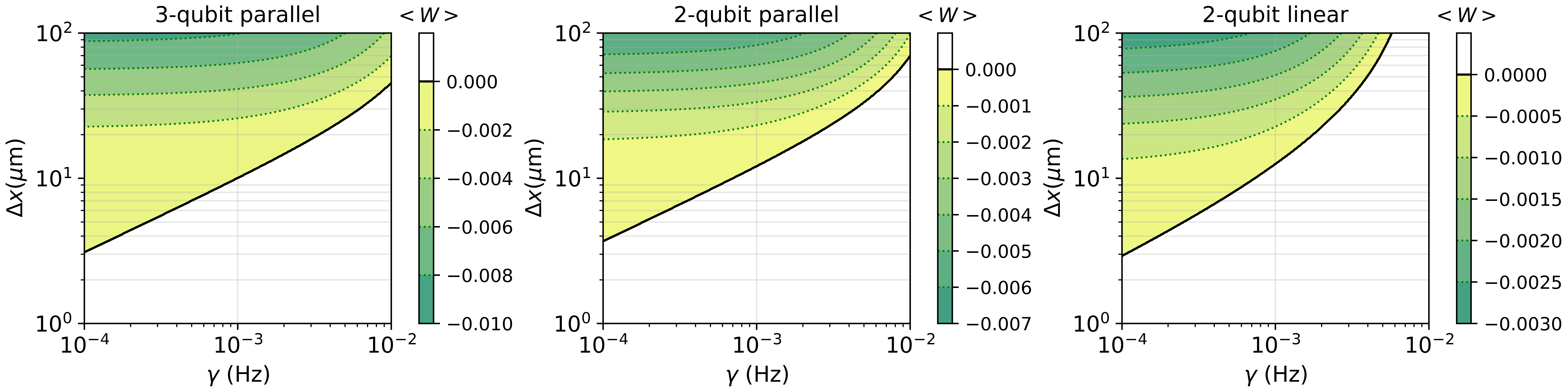}
        \caption{For $m=10^{-15}\,\si{\kg}$}
        \label{fig:par2}
     \end{subfigure}
     \newline
     \begin{subfigure}[b]{\textwidth}
         \centering
         \includegraphics[width=\textwidth]{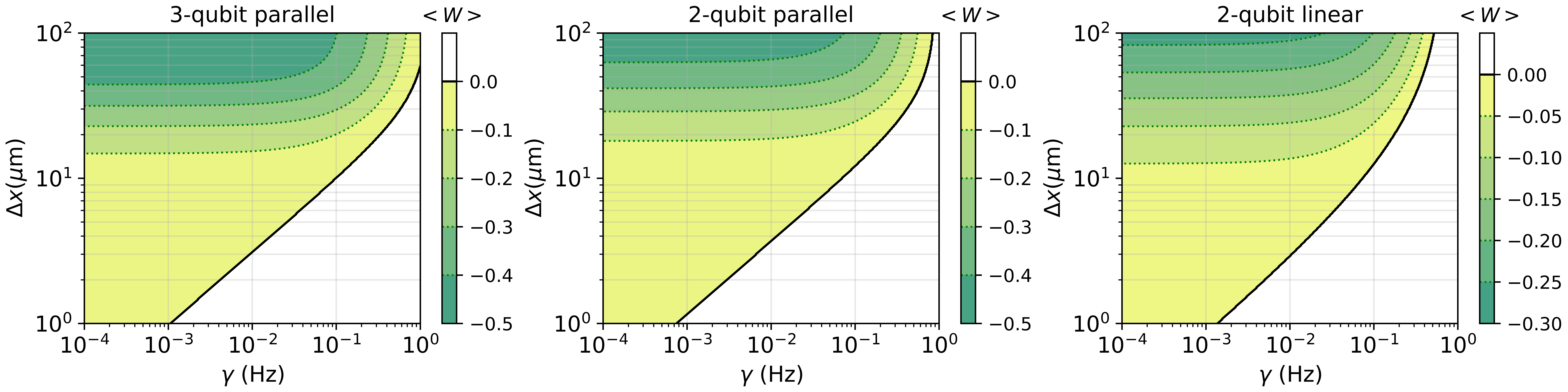}
         \caption{For $m=10^{-14}\,\si{\kg}$}
         \label{fig:3qubit}
     \end{subfigure}
    \caption{For $\tau = 1\,\si{\s}$, and $d_\text{min}=35\,\si{\micro\metre}$.
    The minimum superposition width for the linear 2-qubit setup (eq.~\eqref{eq:witness_deflin}), the parallel setup with 2-qubits (eq.~\eqref{eq:witness_def_par}) and with 3 qubits, as a function of the decoherence rate and with the witness expectation value shown on the colourbar. The top figures are for $m=10^{-15}$ kg, and the bottom figures are for heavier mas $m=10^{-14}$ kg. We can see that heavier mass provides us with an opportunity to create a $\Delta x\sim {\cal O}(1){\rm \mu m}$ for $\gamma\sim 10^{-3}$ Hz for $\langle {\cal W}\rangle > -0.1$. For $\gamma\sim 10^{-1}$ Hz, the superposition size is $\Delta x\sim 11 {\rm \mu m}$ for all three configurations. This parameter space is the most favourable for the detectability of the quantum gravity-induced entanglement.}
    \label{fig:comp}
\end{figure*}

For the parallel setup, the entanglement rate $\omega_1+\omega_2 <0$, thus the witness expectation values will be $\lambda_1$ in Eq.~\eqref{eq:witness_def}:~\footnote{
We can invert eq.~\eqref{eq:witness_def_par} to find the necessary superposition width given the experimental design, i.e. known mass, minimal separation, experimental time and decoherence rate for the desired witness expectation value:
\begin{align}\label{eq:delx}
    (\Delta x)^2 =  &\left(\frac{d_\text{min} G m^2 \tau}{Gm^2\tau+d_\text{min}\hbar\arcsin(\frac{1}{2}e^{-\gamma\tau}[4\langle\mathcal{W}\rangle +1-e^{-2\gamma\tau}])}\right)^2 \nonumber\\
    &\qq{}- d_\text{min}^2
\end{align}
Note, however, that this expression does not take into account the time-dependence of $\Delta x$.}
\begin{equation}\label{eq:witness_def_par}
    \langle \mathcal{W}_\text{par} \rangle = \frac{1}{4} - \frac{1}{4} e^{-\gamma \tau} \left[ e^{-\gamma \tau} - 2 \sin(\frac{\omega_1\tau+\omega_2\tau}{2})\right]
\end{equation}
Note that one can also find an approximate witness expectation value at the level of eq.~\eqref{eq:witness_def_par} by performing an expansion of the trigonometric functions in the limit $\gamma\tau\ll1$ and $(\omega_1+\omega_2)\tau/2\ll1$).
However, since this seems overestimates the negativity of the witness (also partly in the domain we are interested in) we do not consider this approximation here.

Figures~\ref{fig:comp} show the domain where the witness expectation value is negative as a function of the decoherence rate and gives the corresponding superposition width on the y-axis for the parallel setup (the middle plot).
As expected, a higher decoherence rate requires a larger superposition size, and increasing the mass decreases the constraint on the minimal size of the superposition.
At a decoherence rate of $10^{-2}\,\si{\hertz}$ for a witness expectation value $\langle\mathcal{W}\rangle <0$, a mass of $10^{-15}\,\si{\kg}$ would need at least $71\,\si{\micro\metre}$ superposition, while a mass of $10^{-14}\,\si{\kg}$ requires only $4\,\si{\micro\metre}$ superposition.
Figure~\ref{fig:comp} also shows that increasing the mass generates a larger entanglement, thus allowing a smaller superposition size. 
Similarly, reducing the separation $d_\text{min}$ also increases the entanglement generation rate and allows for smaller $\Delta x$ at the same decoherence rate. 

\section{Comparisons}\label{sec:comparison}
Many alternative setups have been proposed in the context of the QGEM experiment.
Here, we highlight the linear setup and the parallel setup. 
From previous analyses, we saw that the linear setup performed noticeably worse~\cite{Tilly:2021qef,Schut:2021svd}, while the 3-qubit parallel setup was shown to be better for higher decoherence rates~\cite{Schut:2021svd}. 
However, the above-mentioned analysis is only true within a certain regime: $\Delta x > d_\text{min}$.
We show that in the regime of experimental parameters that we are entering with improved setups~\footnote{Such as in ref.~\cite{elahi2023probing}, this experimental setup consists of diamagnetically trapped and electromagnetically shielded NVC nanodiamonds; due to the electromagnetic shield the minimal separation between the test masses can be reduced to $35\,\si{\micro\metre}$, see ref.~\cite{Elahi:2024dbb}. Older analyses were done with minimal distances $d_{min}\geq 135\,\si{\micro\metre}$.}
: $\Delta x \sim d_\text{min}$ or even $\Delta x < d_\text{min}$, the linear setup can have some advantages, see~\cite{Schut:2021svd}.
We briefly recap the linear setup and the 3-qubit setup, following ref.~\cite{Schut:2021svd}, and present the necessary experimental parameters as a function of decoherence rate in Fig.~\ref{fig:comp}.

\subsection{Linear setup}

\begin{figure}[t]
\centering
\begin{tikzpicture}
\draw[black, dashed, thick] (-3,0) -- (-2.1,0);
\draw[black, dashed, thick] (-0.5,0) -- (0.4,0);
\draw[black, thick] (-2,-1.3) -- (-2,-1.5) |- (-2,-1.4) -- (-0.5,-1.4) node[pos=0.5, below] {$d_\text{min}$} -| (-0.5,-1.3) -- (-0.5,-1.5);
\draw[black, thick] (-3,-0.7) -- (-3,-0.9) |- (-3,-0.8) -- (-2,-0.8) node[pos=0.5, below] {$\Delta x$} -| (-2,-0.7) -- (-2,-0.9);
\filldraw[black] 
(-3,0) circle (2pt) node[align=center, below] {\hspace{2mm}$\ket{\uparrow}_1$}
(-0.5,0) circle (2pt) node[align=center, below] {\hspace{2mm}$\ket{\uparrow}_2$};
\draw[black]
(-2,0) circle (2pt) node[align=center, below] {\hspace{2mm}$\ket{\downarrow}_1$}
(0.5,0) circle (2pt) node[align=center, below] {\hspace{2mm}$\ket{\downarrow}_2$};
\end{tikzpicture}
\caption{Two neutral test masses of mass $m$, labelled $1$ and $2$, in the linear configuration. The superposition width is $\Delta x$, and the distance between the $\ket{\uparrow}$-states is $d_\text{min} + \Delta x$. A superconducting screen between the two superpositions can mitigate the electromagnetic interactions between the masses, such as dipoles and higher multipoles. However, the gravitational interaction between the two superpositions cannot be screened.}
\label{fig:setup_lin}
\end{figure}
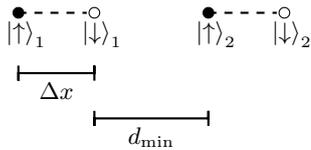

Here, we discuss the `linear' setup as shown in Fig.~\ref{fig:setup_lin}.
The entanglement rates $\omega_1$, $\omega_2$ in eq.~\eqref{WF-1} for the linear configuration are:
\begin{align}
    \omega_1^\text{lin} &= \frac{Gm^2}{d_\text{min}} \frac{1}{\hbar} - \frac{Gm^2}{d_\text{min}+\Delta x} \frac{1}{\hbar} \, , \\
    \omega_2^\text{lin} &= \frac{Gm^2}{d_\text{min}+2\Delta x} \frac{1}{\hbar} - \frac{Gm^2}{d_\text{min}+\Delta x} \frac{1}{\hbar} \, , \label{eq:phases_lin}
\end{align}
The separation between the two superpositions is sometimes expressed as $d$. This paper considers $d = d_\text{min} + \Delta x$ for the linear setup and $d = d_\text{min}$ for the parallel setup.

In the linear setup, the entanglement rate $\omega_1+\omega_2 > 0 $, the value of the witness expectation is thus given by $\lambda_2$ in eq.~\eqref{eq:witness_def}:
\begin{align}
    \langle \mathcal{W}_\text{lin} \rangle = \frac{1}{4} - \frac{1}{4} e^{-\gamma \tau} \left[ e^{-\gamma \tau} + 2 \sin(\frac{\omega_1\tau+\omega_2\tau}{2})\right] \label{eq:witness_deflin}
\end{align}

Figure~\ref{fig:par2} shows that the parallel witness is more robust against decoherence, which is the reason that the parallel setup is often used in literature~\cite{Schut:2021svd,Tilly:2021qef,Chevalier:2020uvv}.
Additionally, the witness expectation value is closer to zero than the parallel setup throughout the plotted ranges, meaning that more measurements are probably needed to find the entanglement.
However, at small decoherence rates, the linear setup allows for slightly smaller superposition sizes for $m=10^{-15}\,\si{\kg}$.
It should be noted that in the presence of the (super)conducting plate separating the two test masses at close proximity, the linear setup will likely suffer more dephasing due to the presence of the plate than the parallel setup~\cite{Schut:2023eux}.

Figure~\ref{fig:3qubit} shows the linear compared to the parallel setup at mass $10^{-14}\,\si{\kg}$.
For this mass, at such small distances, the linear setup requires a smaller $\Delta x$ up to $\gamma\approx10^{-1}\,\si{\hertz}$.
For larger decoherence rates, the parallel setup is favourable.
At $\gamma=10^{-2}\,\si{\hertz}$ the linear setup also requires slightly smaller superposition widths compared to the parallel setup to find $\langle\mathcal{W}\rangle=0$.
At larger separations, $d_\text{min}$, the parallel setup will be favourable, but as we improve the experimental design and can decrease the minimal separation, we can also decrease the superposition size $\Delta x$. 
If we are in the regime $\Delta x\ll d$, the linear setup can become favourable again.
One can see this from the effective entanglement rates $\omega_1+\omega_2$ (see eqs.~\eqref{eq:phases},\eqref{eq:phases_lin}) for both setups:
\begin{align}
    &\abs{\omega_\text{eff}^\text{par}} \approx \frac{Gm^2}{\hbar }\frac{2\Delta x^2}{d^3} \,, \qq{}
    \abs{\omega_\text{eff}^\text{lin}} \approx \frac{Gm^2}{d \hbar} \left[ 1 - \frac{\Delta x}{d} \right] \,, \,\,\, \nonumber
    \\ &\qq{} \text{ for } \Delta x \ll d
\end{align}
The magnitude of the effective entanglement rate is larger for the linear case in this regime.

For $m = 10^{-15}\,\si{\kg}$ we enter this regime only at separations $d_\text{min}\sim\,\si{\micro\metre}$, which are--as of yet--unattainable (because other, electromagnetic, interactions are dominant at this scale, or the magnetic field gradients for these traps have not been considered yet).
This is why Fig.~\ref{fig:par2} shows the parallel setup as being favourable.

\subsection{3-qubit parallel setup}
\begin{figure}[t]
\centering
\begin{tikzpicture}
\draw[black, dashed, thick] (0,-1) -- (0,-0.1);
\draw[black, dashed, thick] (1,-1) -- (1,-0.1);
\draw[black, dashed, thick] (2,-1) -- (2,-0.1);
\draw[black, thick] (0,-1.7) -- (0,-1.9) |- (0,-1.8) -- (1,-1.8) node[pos=0.5, below] {$d_\text{min}$} -| (1,-1.7) -- (1,-1.9);
\draw[black, thick] (-0.6,-1) -- (-0.4,-1) |- (-0.5,-1) -- (-0.5,0) node[pos=0.5, left] {$\Delta x$} -| (-0.6,0) -- (-0.4,0);
\filldraw[black]
(0,-1) circle (2pt) node[align=center, below] {\hspace{2mm}$\ket{\uparrow}_1$}
(1,-1) circle (2pt) node[align=center, below] {\hspace{2mm}$\ket{\uparrow}_2$}
(2,-1) circle (2pt) node[align=center, below] {\hspace{2mm}$\ket{\uparrow}_3$};
\draw[black]
(0,0) circle (2pt) node[align=center, above] {\hspace{2mm}$\ket{\downarrow}_1$}
(1,0) circle (2pt) node[align=center, above] {\hspace{2mm}$\ket{\downarrow}_2$}
(2,0) circle (2pt) node[align=center, above] {\hspace{2mm}$\ket{\downarrow}_3$};
\end{tikzpicture}
\caption{Three neutral test masses of mass $m$, labelled $1$, $2$ and $3$, in the parallel configuration. The superposition width is $\Delta x$, and the distance between neighbouring $\ket{\uparrow}$-states is $d_\text{min}$. It is assumed that the three superpositions are separated by superconducting screens to mitigate the electromagnetic interactions between the masses, such as dipoles and higher multipoles. However, the gravitational interaction between the two superpositions cannot be screened.}
\label{fig:setup_3qgem}
\end{figure}
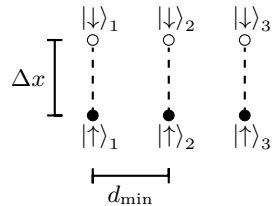

Reference~\cite{Schut:2021svd} investigated the resilience of a 3-qubit setup to decoherence in the context of the QGEM experiment. 
They found that the best 2-dimensional 3-qubit setup is the parallel one, as depicted in Fig.~\ref{fig:setup_3qgem}.

We give a brief recap of the system described in ref.~\cite{Schut:2021svd}.
Instead of Eq.~\eqref{WF-1}, we have the 3-qubit final state resulting from a pure initial state of three spin superpositions:
\begin{equation}\label{eq:psit3}
    \ket{\Psi(t=\tau)}
    =
    \frac{1}{2\sqrt{2}} \sum_{j_1,j_2,j_3=0,1} e^{i \omega_{j_1 j_2 j_3} \tau} \ket{j_1 j_2 j_3},
\end{equation}
where we have replaced the denotation of the spin states $\ket{\uparrow}$ and $\ket{\downarrow}$ by $\ket{0}$ and $\ket{1}$, respectively, for convenience in the equations.
The entanglement rates are defined by:~\cite{Schut:2021svd}
\begin{equation}\label{eq:phase_def}
    \omega_{j_1 j_2 j_3} = \frac{1}{\hbar} \sum_{i,k=1,2,3}^{i<k} \frac{G m^2}{\sqrt{[d(k-i)]^2+[\Delta x (j_i - j_k)]^2}} \, .
\end{equation}
Including the decoherence~\footnote{Analogous to the procedure before: by adding $\bra{i\,j\,k}\rho\ket{i'\,j'\,k'} \rightarrow e^{-\gamma \tau (3 - \delta_{i,i'} - \delta_{j,j'} - \delta_{k,k'})}$ ), we add an exponential decay to the off-diagonal terms in the density matrix.} we can find the witness expectation value numerically by taking the smallest eigenvalue of the partial transpose of the density matrix derived from eq.~\eqref{eq:psit3} via $\rho = \ket{\Psi}\bra{\Psi}$.

The witness for the 3-qubit case is a sufficient but not necessary condition for entanglement. 
We do not give the expression here analytically, but instead use the partial transpose or the density matrix to find the lowest eigenvalue numerically.
Note that the 3-qubit witness has a different and larger basis, which means that more repeated measurements will be necessary to find the entanglement (up to some $\gamma$).
Note also that the bi-partition is chosen to be $13|2$, meaning we take the partial transpose over system $2$. Other choices of bi-partition will give an entanglement, but a smaller entanglement generation rate comparable to the 2-qubit setup.
For more details, we refer to ref.~\cite{Schut:2021svd}.

Figure~\ref{fig:comp} shows contour plots for the witness expectation value as a function of decoherence rate and superposition size.
As found in ref.~\cite{Schut:2021svd}, the 3-qubit setup is more resilient to decoherence, compared to the 2-qubit parallel case; this is confirmed in figs.~\ref{fig:comp} for the experimental parameters one would expect in the shielded and trapped chip setup of ref.~\cite{elahi2023probing}.
Figure~\ref{fig:par2} shows that even at $10^{-2}\,\si{\hertz}$, the gain from the 3-qubit setup is noticeable: one would require $\sim45\,\si{\micro\metre}$ superposition, compared to a $\sim70\,\si{\micro\metre}$ superposition for the 2-qubit parallel case~\footnote{Note that comparing the witness expectation value between the 2-and 3-qubit case is tricky because this is basis-dependent and the witnesses are given in different bases.}.
Extending the QGEM protocol to a 3-qubit setup should not be too experimentally challenging in the current setups, which include trapping and shielding via a chip-based setup~\cite{elahi2023probing}, as it simply requires the addition of a second chip with wires on only one side.
The magnetic field that levitates the second (middle) qubit will, however, interact with the additional chip/conducting plate, which might cause a slight change in the trapping minimum's position.
This can be countered by changing the bias magnetic field by adjusting the current in the wire on the chip.

For $m=10^{-15}\,\si{\kg}$ (Fig.~\ref{fig:par2}), it would seem experimentally realisable and also favourable to go for the 3-qubit parallel setup.
For $m=10^{-14}\,\si{\kg}$ and $d_\text{min} = 35\,\si{\micro\metre}$, the 3-qubit setup is also most resilient against decoherence; see Fig.~\ref{fig:3qubit}.
However, at smaller $\gamma$, the linear setup will require the smallest $\Delta x$.
In the expected decoherence regime: $\gamma\in[1-^{-4},10^{-1}]\,\si{\hertz}$, the difference between the three setups is not too noticeable. 
This was not found in ref.~\cite{Schut:2021svd} due to the current parameters not being available yet during that work, as they are a result of later works~\cite{Schut:2023eux,vandeKamp:2020rqh,elahi2023probing,Schut:2023hsy}.
The difference between the three cases in terms of resilience against decoherence in Fig.~\ref{fig:3qubit} is not as relevant in Fig.~\ref{fig:par2}, due to the expected decoherence rate being $<0.1\,\si{\hertz}$.
The drawback of the 3-qubit setup, having an exponential increase in a number of terms in the witness, would mean that for $m=10^{-14}\,\si{\kg}$ the parallel and linear setups seem the most favourable, however, we have mentioned before that the linear setup might suffer more from dephasing due to interaction with the plate, compared to any-number parallel setup~\cite{Schut:2023eux}.
If the additional dephasing in the 2-qubit linear setup cannot be mitigated, we would expect that the parallel setup is still favourable at $m=10^{-14}\,\si{\kilogram}$.

Alternatively, one could also consider a 3-dimensional parallel setup such that all superposition states $\ket{0}$ ($\ket{\uparrow}$) form an equilateral triangle with edge length $d_\text{min}$ and all superposition states $\ket{1}$ ($\ket{\downarrow}$) form the same triangle at a distance $\Delta x$ in the direction of the triangle plane.
In this case, the phase would be given by:
\begin{equation}\label{eq:phase_def}
    \omega_{j_1 j_2 j_3} = \frac{1}{\hbar} \sum_{i,k=1,2,3}^{i<k} \frac{G m^2}{\sqrt{d^2+[\Delta x (j_i - j_k)]^2}} \, .
\end{equation}
Following the arguments presented in ref.~\cite{Schut:2021svd}, which compared several 2-dimensional 3-qubit setups, one might expect this setup to be favourable regarding witness expectation value because of the reduced average distance between neighbouring test masses. 
However, due to the favourable tracing of the second subsystem in the 2-dimensional 3-qubit setup of Fig.~\ref{fig:setup_3qgem}, there is not much to gain by extending the 3-qubit parallel setup to three dimensions.
The average distance between the bipartite subsystems stays the same.

\section{Results \& Conclusion}\label{sec:conclusion}
This paper gives a brief analysis of the 2-qubit parallel and linear setups, as well as the 3-qubit parallel setup, in terms of experimental parameters in the context of the trapped and shielded setup such as e.g. proposed in ref.~\cite{Schut:2023hsy,elahi2023probing}.
As found in previous works~\cite{Schut:2021svd,Tilly:2021qef,Chevalier:2020uvv}, the parallel setups have a negative witness expectation value up to higher decoherence rates for $m=10^{-15}\,\si{\kg}$.
For this mass, the 3-qubit parallel setup also clearly shows to be more resilient to decoherence than the 2-qubit setup and allows for smaller spatial superposition sizes.
For a larger mass, $m=10^{-14}\,\si{\kg}$, the difference between the three setups is small for $\gamma\in[10^{-4},10^{-1}]\,\si{\hertz}$.

Although previous papers found the necessary $\Delta x$ with an approximated witness expectation value, we have here considered the exact expression by taking the minimal eigenvalue of the partial transpose of the density matrix; this gives a slightly more pessimistic view in terms of the necessary experimental parameters~\footnote{For example, a mass $m=10^{-15}\,\si{\kilogram}$ at $d_\text{min}=21\,\si{\micro\metre}$ in the parallel setup requires at least $\Delta x = 24\,\si{\micro\metre}$ superposition instead of the $23\,\si{\micro\metre}$ given in ref.~\cite{Schut:2023hsy} for $\gamma=10^{-2}\,\si{\hertz}$. However, the difference becomes more noticeable at higher decoherence rates.}.
We have additionally updated previous analyses for three different setups (2-qubit parallel, 2-qubit linear, and 3-qubit parallel) by considering the most recent experimental setup, which uses trapping and shielding to lessen the separation $d_\text{min}$ and mitigate the electromagnetic interactions between the superpositions, see ref.~\cite{Elahi:2024dbb}.

This smaller minimal separation opens the door to a new regime in the $10^{-14}\,\si{\kilogram}$ case, where we find $\Delta x \sim d_\text{min}$, or $\Delta x<d_\text{min}$. 
In this regime, we find that the linear setup becomes slightly favourable, in terms of allowing a smaller superposition size for $\gamma<10^{-1}\,\si{\hertz}$.
However, the witness expectation value of the linear setup is less negative than that of the parallel setup, which could potentially indicate the necessity of more repeated measurements.
Additionally, it is expected that the linear setup will suffer more from dephasing in a shielded setup than any parallel setup.
If this can be mitigated and if the separation can be further reduced, the linear setup could become favourable, especially if it allows for smaller spatial superposition widths.

The creation of large spatial superpositions in masses is extremely challenging~\cite{Wan16_GM,Scala13_GM,
Bose:2017nin,Pedernales:2020nmf,Marshman:2021wyk,Marshman:2018upe,Zhou:2022epb,Zhou:2022frl,Zhou:2022jug,Zhou:2024voj,Braccini:2024fey}.
While increasing the mass and decreasing the decoherence rate~\footnote{
Here, we have not taken into account that the decoherence rate scales with radius, nor that it scales with the superposition size; we have taken a generic parameter $\gamma\in[10^{-4},10^{-1}]\,\si{\hertz}$.
} are ways of lowering the requirement on the minimal necessary superposition size, these are experimentally challenging; choosing the right setup will help relax these challenges at least a little bit. 

Our analysis suggests the 2-qubit parallel setup is still best for mass $m=10^{-14}$~kg. 
It can be realised experimentally on a chip, see~\cite{Elahi:2024dbb}, and it is comparable to all other setups in terms of witness for $\gamma<10^{-1}\,\si{\hertz}$ and for $d_{min}=35 \rm {\mu m}$. 
The superposition size $\Delta x\sim {\cal O}(1)$ for $\gamma = 10^{-3}$ Hz. 
If the decoherence rate increases to $\gamma =10^{-2}$ Hz, then $\Delta x\sim 3 {\rm \mu m}$, and for $\gamma =10^{-1}$Hz, $\Delta x \sim 11 {\rm \mu m}$.

These numbers are based on the PPT witness; they open up the possibility of future investigations using different kinds of witnesses in the QGEM protocol.
Since the total entanglement phase depends on the specific protocol of creating the spatial superpositions, for which many variations have been proposed, we have taken $\Delta x$ to be constant during a time $\tau$ and have not considered the entanglement phase picked up during its time-dependence.
Nevertheless, since during the creation/annihilation, the $\Delta x\ll d$, the entanglement rate will be low during these times. Although including the time dependence will change the witness expectation value, albeit with a very small effect, the comparison we make between the different setups will hold. \newline


\noindent {\it Acknowledgements:}
AM’s research is funded by the Gordon and Betty
Moore Foundation through Grant GBMF12328, DOI
10.37807/GBMF12328.)

\bibliography{casimir.bib} 
\bibliographystyle{ieeetr}

\end{document}